\documentclass[aps,prl,twocolumn,showpacs,superscriptaddress,showpacs,
preprintnumbers]{revtex4-1}

\usepackage{xurl}
\usepackage{graphicx}  
\usepackage{dcolumn}   
\usepackage{bm}        
\usepackage{amsmath}

\usepackage{amssymb}   
\usepackage{xcolor}    
\usepackage{latexsym}
\usepackage{amsfonts}

\definecolor{purple}{rgb}{0.5,0,0.5}
\definecolor{blue}{rgb}{0.0,0,0.9}
\definecolor{prdblue}{rgb}{0.133,0.118,0.498}
\usepackage[colorlinks=true, pdfstartview=FitV, linkcolor=prdblue, citecolor= prdblue, urlcolor=prdblue]{hyperref}

\begin{document}

\title{Polarized Structure Function \boldmath{$\sigma_{LT'}$} from $\pi^0 p$ Electroproduction Data in the Resonance Region at $0.4$~GeV$^2 < Q^2 < 1.0$~GeV$^2$}

\newcommand*{\ANL}{Argonne National Laboratory, Argonne, Illinois 60439}
\newcommand*{\ANLindex}{1}
\affiliation{\ANL}
\newcommand*{\CSUDH}{California State University, Dominguez Hills, Carson, CA 90747}
\newcommand*{\CSUDHindex}{2}
\affiliation{\CSUDH}
\newcommand*{\CMU}{Carnegie Mellon University, Pittsburgh, Pennsylvania 15213}
\newcommand*{\CMUindex}{3}
\affiliation{\CMU}
\newcommand*{\CUA}{Catholic University of America, Washington, D.C. 20064}
\newcommand*{\CUAindex}{4}
\affiliation{\CUA}
\newcommand*{\SACLAY}{IRFU, CEA, Universit\'{e} Paris-Saclay, F-91191 Gif-sur-Yvette, France}
\newcommand*{\SACLAYindex}{5}
\affiliation{\SACLAY}
\newcommand*{\CNU}{Christopher Newport University, Newport News, Virginia 23606}
\newcommand*{\CNUindex}{6}
\affiliation{\CNU}
\newcommand*{\UCONN}{University of Connecticut, Storrs, Connecticut 06269}
\newcommand*{\UCONNindex}{7}
\affiliation{\UCONN}
\newcommand*{\DUQUESNE}{Duquesne University, 600 Forbes Avenue, Pittsburgh, PA 15282 }
\newcommand*{\DUQUESNEindex}{8}
\affiliation{\DUQUESNE}
\newcommand*{\FU}{Fairfield University, Fairfield CT 06824}
\newcommand*{\FUindex}{9}
\affiliation{\FU}
\newcommand*{\FERRARAU}{Universita' di Ferrara , 44121 Ferrara, Italy}
\newcommand*{\FERRARAUindex}{10}
\affiliation{\FERRARAU}
\newcommand*{\FIU}{Florida International University, Miami, Florida 33199}
\newcommand*{\FIUindex}{11}
\affiliation{\FIU}
\newcommand*{\FSU}{Florida State University, Tallahassee, Florida 32306}
\newcommand*{\FSUindex}{12}
\affiliation{\FSU}
\newcommand*{\GWUI}{The George Washington University, Washington, DC 20052}
\newcommand*{\GWUIindex}{13}
\affiliation{\GWUI}
\newcommand*{\INFNFE}{INFN, Sezione di Ferrara, 44100 Ferrara, Italy}
\newcommand*{\INFNFEindex}{14}
\affiliation{\INFNFE}
\newcommand*{\INFNFR}{INFN, Laboratori Nazionali di Frascati, 00044 Frascati, Italy}
\newcommand*{\INFNFRindex}{15}
\affiliation{\INFNFR}
\newcommand*{\INFNGE}{INFN, Sezione di Genova, 16146 Genova, Italy}
\newcommand*{\INFNGEindex}{16}
\affiliation{\INFNGE}
\newcommand*{\INFNRO}{INFN, Sezione di Roma Tor Vergata, 00133 Rome, Italy}
\newcommand*{\INFNROindex}{17}
\affiliation{\INFNRO}
\newcommand*{\INFNTUR}{INFN, Sezione di Torino, 10125 Torino, Italy}
\newcommand*{\INFNTURindex}{18}
\affiliation{\INFNTUR}
\newcommand*{\INFNCAT}{INFN, Sezione di Catania, 95123 Catania, Italy}
\newcommand*{\INFNCATindex}{19}
\affiliation{\INFNCAT}
\newcommand*{\INFNPAV}{INFN, Sezione di Pavia, 27100 Pavia, Italy}
\newcommand*{\INFNPAVindex}{20}
\affiliation{\INFNPAV}
\newcommand*{\ORSAY}{Universit'{e} Paris-Saclay, CNRS/IN2P3, IJCLab, 91405 Orsay, France}
\newcommand*{\ORSAYindex}{21}
\affiliation{\ORSAY}
\newcommand*{\Juelich}{Institute fur Kernphysik (Juelich), Juelich, Germany}
\newcommand*{\Juelichindex}{22}
\affiliation{\Juelich}
\newcommand*{\KNU}{Kyungpook National University, Daegu 41566, Republic of Korea}
\newcommand*{\KNUindex}{23}
\affiliation{\KNU}
\newcommand*{\LAMAR}{Lamar University, 4400 MLK Blvd, PO Box 10046, Beaumont, Texas 77710}
\newcommand*{\LAMARindex}{24}
\affiliation{\LAMAR}
\newcommand*{\MISS}{Mississippi State University, Mississippi State, MS 39762-5167}
\newcommand*{\MISSindex}{25}
\affiliation{\MISS}
\newcommand*{\ITEP}{National Research Centre Kurchatov Institute - ITEP, Moscow, 117259, Russia}
\newcommand*{\ITEPindex}{26}
\affiliation{\ITEP}
\newcommand*{\UNH}{University of New Hampshire, Durham, New Hampshire 03824-3568}
\newcommand*{\UNHindex}{27}
\affiliation{\UNH}
\newcommand*{\NMSU}{New Mexico State University, PO Box 30001, Las Cruces, NM 88003, USA}
\newcommand*{\NMSUindex}{28}
\affiliation{\NMSU}
\newcommand*{\NSU}{Norfolk State University, Norfolk, Virginia 23504}
\newcommand*{\NSUindex}{29}
\affiliation{\NSU}
\newcommand*{\OHIOU}{Ohio University, Athens, Ohio  45701}
\newcommand*{\OHIOUindex}{30}
\affiliation{\OHIOU}
\newcommand*{\ODU}{Old Dominion University, Norfolk, Virginia 23529}
\newcommand*{\ODUindex}{31}
\affiliation{\ODU}
\newcommand*{\JLUGiessen}{II Physikalisches Institut der Universitaet Giessen, 35392 Giessen, Germany}
\newcommand*{\JLUGiessenindex}{32}
\affiliation{\JLUGiessen}
\newcommand*{\RPI}{Rensselaer Polytechnic Institute, Troy, New York 12180-3590}
\newcommand*{\RPIindex}{33}
\affiliation{\RPI}
\newcommand*{\URICH}{University of Richmond, Richmond, Virginia 23173}
\newcommand*{\URICHindex}{34}
\affiliation{\URICH}
\newcommand*{\ROMAII}{Universita' di Roma Tor Vergata, 00133 Rome Italy}
\newcommand*{\ROMAIIindex}{35}
\affiliation{\ROMAII}
\newcommand*{\MSU}{Skobeltsyn Institute of Nuclear Physics, Lomonosov Moscow State University, 119234 Moscow, Russia}
\newcommand*{\MSUindex}{36}
\affiliation{\MSU}
\newcommand*{\SCAROLINA}{University of South Carolina, Columbia, South Carolina 29208}
\newcommand*{\SCAROLINAindex}{37}
\affiliation{\SCAROLINA}
\newcommand*{\TEMPLE}{Temple University,  Philadelphia, PA 19122 }
\newcommand*{\TEMPLEindex}{38}
\affiliation{\TEMPLE}
\newcommand*{\JLAB}{Thomas Jefferson National Accelerator Facility, Newport News, Virginia 23606}
\newcommand*{\JLABindex}{39}
\affiliation{\JLAB}
\newcommand*{\UTFSM}{Universidad T\'{e}cnica Federico Santa Mar\'{i}a, Casilla 110-V Valpara\'{i}so, Chile}
\newcommand*{\UTFSMindex}{40}
\affiliation{\UTFSM}
\newcommand*{\BRESCIA}{Universit\`{a} degli Studi di Brescia, 25123 Brescia, Italy}
\newcommand*{\BRESCIAindex}{41}
\affiliation{\BRESCIA}
\newcommand*{\MESSU}{Universit`{a} degli Studi di Messina, 98166 Messina, Italy}
\newcommand*{\MESSUindex}{42}
\affiliation{\MESSU}
\newcommand*{\GLASGOW}{University of Glasgow, Glasgow G12 8QQ, United Kingdom}
\newcommand*{\GLASGOWindex}{43}
\affiliation{\GLASGOW}
\newcommand*{\YORK}{University of York, York YO10 5DD, United Kingdom}
\newcommand*{\YORKindex}{44}
\affiliation{\YORK}
\newcommand*{\VT}{Virginia Tech, Blacksburg, Virginia   24061-0435}
\newcommand*{\VTindex}{45}
\affiliation{\VT}
\newcommand*{\VIRGINIA}{University of Virginia, Charlottesville, Virginia 22901}
\newcommand*{\VIRGINIAindex}{46}
\affiliation{\VIRGINIA}
\newcommand*{\WM}{College of William and Mary, Williamsburg, Virginia 23187-8795}
\newcommand*{\WMindex}{47}
\affiliation{\WM}
\newcommand*{\YEREVAN}{Yerevan Physics Institute, 375036 Yerevan, Armenia}
\newcommand*{\YEREVANindex}{48}
\affiliation{\YEREVAN}
\newcommand*{\RIVER}{University of California, Riverside, California, 92521, USA}
\newcommand*{\RIVERindex}{49}
\affiliation{\RIVER}
\newcommand*{\CANISIUS}{Canisius College, Buffalo, NY}
\newcommand*{\CANISIUSindex}{50}
\affiliation{\CANISIUS}
\newcommand*{\LLNL}{Lawrence Livermore National Laboratory, Livermore, CA, 94550}
\newcommand*{\LLNLindex}{51}
\affiliation{\LLNL}

\newcommand*{\NOWISU}{Idaho State University, Pocatello, Idaho 83209}

\affiliation{\VIRGINIA}

\author{E.L.~Isupov}
\affiliation{\MSU}
\affiliation{\UCONN}
\author {V.D.~Burkert} 
\affiliation{\JLAB}
\author{A.A.~Golubenko}
\affiliation{\MSU}
\author{K.~Joo}
\affiliation{\UCONN}
\author{N.S.~Markov}
\affiliation{\JLAB}
\affiliation{\UCONN}
\author{V.I.~Mokeev}
\affiliation{\JLAB}
\author{L.C.~Smith}
\affiliation{\VIRGINIA}
\author {W.R. Armstrong} 
\affiliation{\ANL}
\author {H.~Atac} 
\affiliation{\TEMPLE}
\author {H.~Avakian}
\affiliation{\JLAB}
% ...

\collaboration{The CLAS Collaboration}
\noaffiliation



\begin{abstract}
The first results on the $\sigma_{LT'}$ structure function in exclusive $\pi^0p$ electroproduction at invariant masses of the
final state of 1.5 GeV $<$ $W$ $<$ 1.8 GeV and in the range of photon virtualities 0.4~GeV$^2 < Q^2 < 1.0$~GeV$^2$ were
obtained from data on beam spin asymmetries and differential cross sections measured with the CLAS detector at Jefferson Lab. 
The Legendre moments determined from the $\sigma_{LT'}$ structure function have demonstrated  sensitivity to the contributions 
from the nucleon resonances in the second and third resonance regions. These new data on the beam spin asymmetries in $\pi^0p$
electroproduction extend the opportunities for the extraction of the nucleon resonance electroexcitation amplitudes in the mass
range above 1.6 GeV.    
\end{abstract}

\maketitle


Studies of $\pi N$ electroproduction are an effective tool for the exploration of  nucleon resonance structure
\cite{Mokeev20,aznaryanBurkert,BurkertRoberts,sqcd19,aznauryan}. The CEBAF Large Acceptance Spectrometer (CLAS) at Jefferson Lab
has provided most of the  available information on these electroproduction channels at invariant masses of the final state hadrons
$W < 1.8$~GeV and at photon virtualities $Q^2 < 5.0$~GeV$^2$~\cite{JooSmith:2002,Ungaro:2006,Bis08,parkAznauryan,markov20}. 
The available data allow us to determine the nucleon resonance electroexcitation amplitudes ({\it i.e.} the $\gamma_{v}pN^*$
electro\-couplings) for most resonances over this kinematic range~\cite{Mokeev20,aznauryan,aznaryanBurkert,parkAznauryan}. These results 
allow us to evaluate the resonant contribution to inclusive electron scattering with the $\gamma_{v}pN^*$ electrocouplings available from 
the experimental data, thereby paving the way to gain insight into the parton distributions in the ground state nucleon within the resonance 
excitation region~\cite{blin19,blin21}. High-level analyses of the results on the $\gamma_{v}pN^*$ electro\-couplings have revealed the 
structure of nucleon resonances as a complex interplay between the inner core of three dressed quarks and an external meson baryon cloud
\cite{Mokeev20,aznaryanBurkert,BurkertRoberts}, shed light on the emergence of hadron mass~\cite{Mokeev20,sqcd19,cdr19,Seg19},
and allow for the modeling of $N^*$ structure within different quark models~\cite{aznuryan14,aznuryan17,obukh19,lub20,ram141,santopinto1}. 

CLAS studies of $\pi^+n$ electroproduction~\cite{parkAznauryan} have provided the $\gamma_{v}pN^*$ electrocouplings for the $N(1675)5/2^-$, 
$N(1680)5/2^+$, and $N(1710)1/2^+$ resonances~\cite{aznuryan14}. For a complete isospin analysis, it is important to explore both the $\pi^+n$ 
and $\pi^0p$ channels. Recently, new CLAS results on the differential $\pi^0p$ electroproduction cross sections have become available for 
$W < 1.8$~GeV and 0.4~GeV$^2 < Q^2 < 1.0$~GeV$^2$~\cite{markov20}. 
However, the data on $\pi N$ electroproduction at $W > 1.6$~GeV and $Q^2 < 1.0$~GeV$^2$ consist mostly of measurements with an unpolarized 
electron beam and an unpolarized proton target.

Measurement of the beam spin asymmetry (BSA) and the related $\sigma_{LT'}$ structure function can provide important constraints on the 
extraction of the $\gamma_{v}pN^*$ electro\-couplings when combined with the differential cross sections. The $\sigma_{LT'}$ structure function 
determines the imaginary part of bilinear products between longitudinal and transverse amplitudes. Small contributions from the imaginary part
of the longitudinal resonance amplitudes are amplified in their interference with the real part of the non-resonant contributions, making the 
BSAs an important observable for extraction of the longitudinal $S_{1/2}$ $\gamma_{v}pN^*$ electrocouplings. Previous studies of BSAs in both 
the $\pi^+n$ and $\pi^0p$ channels~\cite{Joo05,Joo04,Bis03,Joo03} were focused in the range of $W < 1.5$~GeV. They demonstrated a substantial 
impact of the BSA data on the extraction of the $\Delta(1232)3/2^+$ and $N(1440)1/2^+$ $S_{1/2}$ electrocouplings published in 
Ref.~\cite{aznauryan}. 

In this Letter, we present new measurements of the BSAs and $\sigma_{LT'}$ structure function from the CLAS $\pi^0p$ electroproduction data 
with a major focus on the exploration of the second and third resonance regions. The data reported here cover the kinematic area of 
1.5~GeV $< W < 1.8$~GeV and 0.4~GeV$^2 < Q^2 < 1.0$~GeV$^2$ with the full angular range in the center-of-mass (CM) frame. 

The data were taken during the e1e run period with the CLAS detector~\cite{CLASDetector}. A longitudinally polarized electron beam of 2.036-GeV 
energy and 10-nA nominal current was incident on a 2-cm-long liquid-hydrogen ($LH_2$) target. The beam polarization determined from M${\o}$ller 
polarimeter measurements was 78.9$\pm$2.8(stat)$\pm$3.0(syst)\%. 
All details on particle identification, event selection, and the related systematic uncertainties are available in our previous publication on 
$\pi^0 p$ cross sections and exclusive structure functions~\cite{markov20}. 

The kinematics of the $ep \to$ $e' \pi^0 p$ reaction can be fully described by $Q^2$, $W$ and the final-state pion polar $\theta_{\pi}$ and 
azimuthal $\phi_{\pi}$ emission angles in the CM frame, where $\phi_{\pi}$ is defined relative to the electron scattering plane.
The exclusive $\pi^0p$ events were identified from the missing mass squared $M^2_X$ distributions in the reaction $ep \to e' p X$ after 
application of kinematic cuts to eliminate Bethe-Heitler backgrounds~\cite{markov20}. The exclusivity cuts over $M^2_X$ were applied in 
3-dimensional bins of $(Q^2,W,\cos\theta_{\pi})$.   
The selected $\pi^0p$ events were binned using a $(Q^2,W,\cos\theta_{\pi},\phi_{\pi})$ grid consisting of 2 bins over $Q^2$: from 
0.4~GeV$^2$ to 0.6~GeV$^2$ and from 0.6~GeV$^2$ to 1.0~GeV$^2$, 28 bins over $W$, 10 bins over $\cos\theta_{\pi}$, and 12 bins over $\phi_{\pi}$ of 
equal sizes.



\begin{figure}[htb]
\includegraphics[scale=0.3]{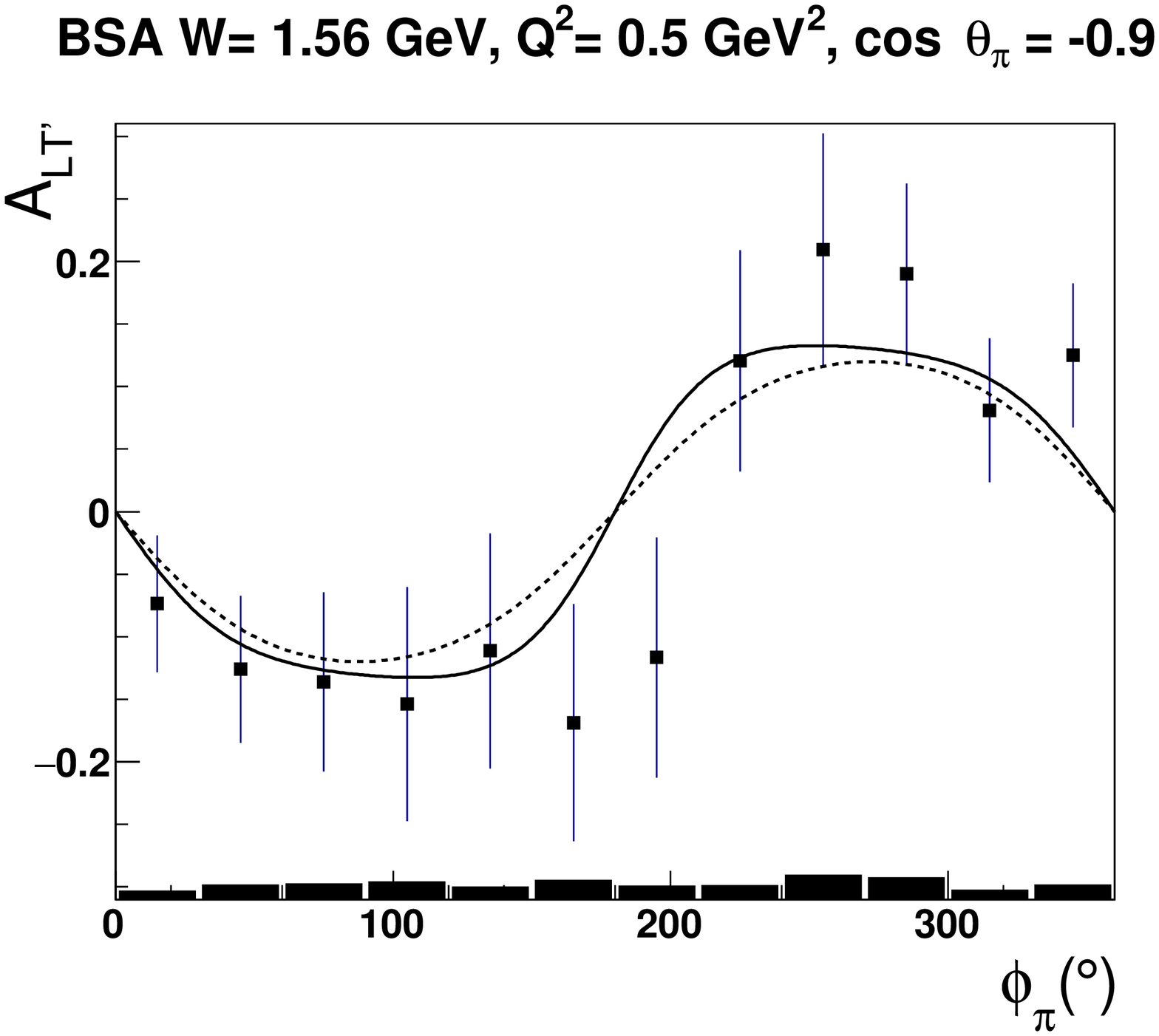}
\includegraphics[scale=0.3]{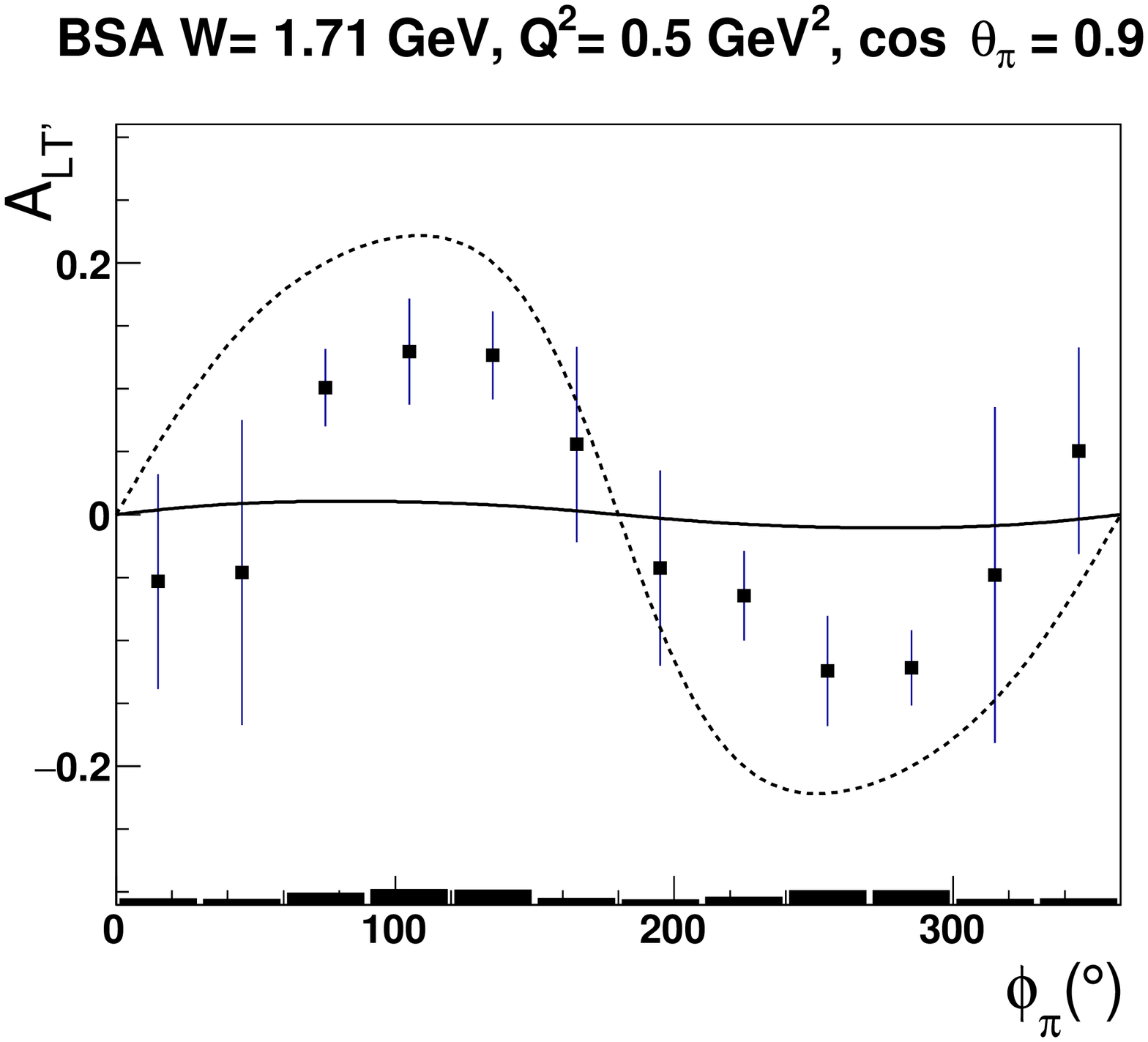}
\caption{BSA as a function of the CM pion azimuthal angle $\phi_{\pi}$ for the $ep \to e'\pi^0p$ reaction at $W$=1.56~GeV, $Q^2=0.5$~GeV$^2$,
$\cos\theta_{\pi}=-0.9$ (top) and at $W$=1.71~GeV, $Q^2=0.5$~GeV$^2$, $\cos\theta_{\pi}=0.9$ (bottom). The MAID model~\cite{MAID} expectation with 
the $\gamma_{v}pN^*$ electrocouplings from MAID~\cite{Tiator:2011pw} and the CLAS analyses~\cite{isupov-web,blin19,mok20p} are shown by the solid and 
dotted lines, respectively.}
\label{fig:BSA08}
\end{figure}

The measured asymmetry $A_{LT'}(Q^2,W,\cos\theta_{\pi},\phi_{\pi})$ was obtained from the event yields produced by the incoming 
electrons of positive and negative helicities, $N_\pi^{+}(Q^2,W,\cos\theta_{\pi},\phi_{\pi})$ and $N_\pi^{-} (Q^2,W,\cos\theta_{\pi},\phi_{\pi})$:
\begin{eqnarray}
A_{LT'} =  \frac{1}{P_e} \frac{(N_\pi^{+} - N_\pi^-)}{(N_\pi^{+} + N_\pi^{-})}, 
\label{eq:altp_m}
\end{eqnarray}
where $P_e$ is the electron beam polarization. Representative examples of the BSA in the $ep \to e' \pi^0 p$ reaction at $W$ values in the 
second and third resonance regions are shown in Fig.~\ref{fig:BSA08}. The full set of BSA data from our measurement is available in the CLAS 
Physics Database~\cite{clasdb}. The data in Fig.~\ref{fig:BSA08} are compared with the MAID model~\cite{MAID} predictions for the BSA computed with
the $\gamma_{v}pN^*$ electrocouplings from CLAS analyses of $\pi N$ and $\pi^+\pi^-p$ data ~\cite{aznauryan,isupov-web,blin19,mok20p,aznuryan14,mok20p} 
and MAID analyses \cite{Tiator:2011pw} of the $\pi N$ electroproduction data. 
The BSAs predicted with the resonance electrocouplings from both the CLAS and MAID analyses are consistent in the second resonance region and 
reasonably reproduce the BSA data (see Fig.~\ref{fig:BSA08} (top)), demonstrating consistency of the MAID and CLAS analysis results on resonance
electroexcitation in the second resonance region. 
In the third resonance region the BSAs computed with the CLAS and MAID results on the $\gamma_{v}pN^*$ electrocouplings
are substantially different.
Our results will be essential to improve the knowledge on the $\gamma_{v}pN^*$ electrocouplings of the resonances in the third resonance region. 


The $\gamma_{v}p \to \pi^0p$ virtual photon differential cross sections  $\frac{d\sigma}{d\Omega_{\pi}}(Q^2,W,\cos\theta_{\pi},\phi_{\pi})$ 
for an electron beam of helicity $h$ ($h=\pm 1$) off unpolarized protons depends on five structure functions. The transverse
$\sigma_{T}(Q^2,W,\cos\theta_{\pi})$, the longitudinal $\sigma_{L}(Q^2,W,\cos\theta_{\pi})$, the transverse-transverse 
$\sigma_{TT}(Q^2,W,\cos\theta_{\pi}) $, and the longitudinal-transverse $\sigma_{LT}(Q^2,W,\cos\theta_{\pi})$ structure functions describe the 
helicity-independent part of the differential cross section, while the part proportional to the electron beam helicity $h$ is described by the
$\sigma_{LT'}(Q^2,W,\cos\theta_{\pi})$ structure function~\cite{MAID}: 
\begin{eqnarray}
\frac{d\sigma}{d\Omega_{\pi}} &=& \frac{p^*_{\pi}}{k_{\gamma}^*} (\sigma_{0} +
h\sqrt{2\epsilon_L(1-\epsilon)}\,\sigma_{LT'}\,\sin\,\theta_{\pi}\,\sin\,\phi_{\pi}),  \nonumber 
\\
\sigma_{0} &=& \sigma_T+\epsilon_L\sigma_L+\epsilon\,\sigma_{TT}\,\sin^2\theta_{\pi}\,\cos\,2\phi_{\pi} \nonumber \\
~&+&
\sqrt{2\epsilon_L(1+\epsilon)}\,\sigma_{LT}\,\sin\,\theta_{\pi}\,\cos\,\phi_{\pi},
\label{eq:str}
\end{eqnarray}
where $p_{\pi}^*$ is the magnitude of the $\pi^0$ momentum in the CM frame.
The commonly used real photon equivalent energy $k_{\gamma}^*$ and the virtual photon polarization parameters $\epsilon$ and $\epsilon_L$ are described 
in Ref.~\cite{MAID}. 

\begin{figure}[htb]

\includegraphics[scale=0.5]{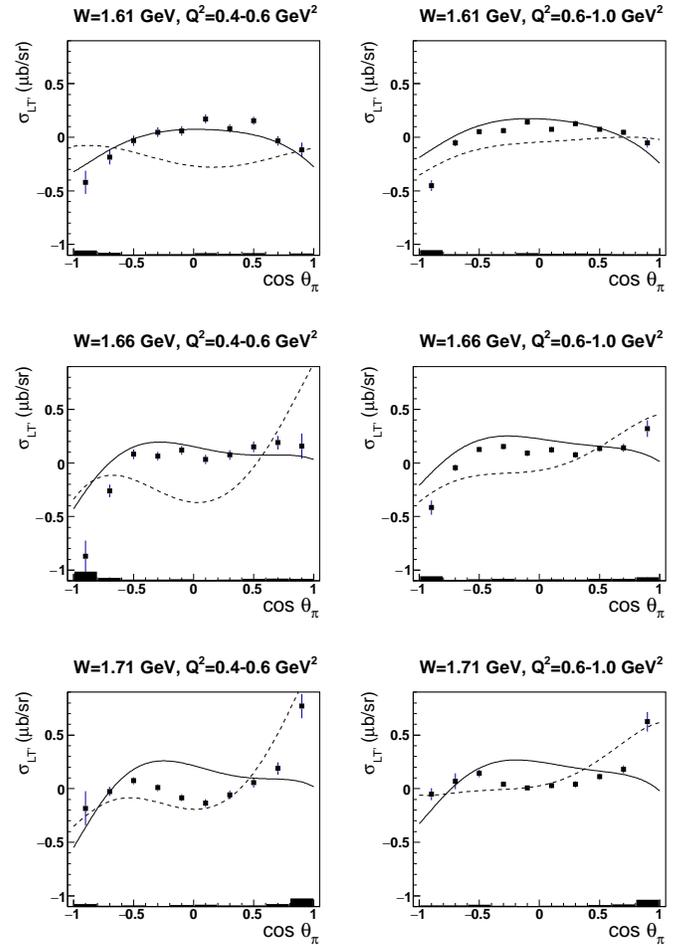}
\caption{The structure function $\sigma_{LT'}$ for $\pi^0p$ electroproduction off protons at $W > 1.6$~GeV,  $Q^2$ from 0.4~GeV$^2$ to 0.6~GeV$^2$ (left panel) and from 0.6~GeV$^2$ to 1.0~GeV$^2$ (right panel). 
The lines represent the evaluations within the MAID reaction model~\cite{MAID} with the $\gamma_{v}pN^*$ electrocouplings from the $\pi N$ 
electroproduction analysis~\cite{Tiator:2011pw} (solid line) and from the CLAS exclusive meson electroproduction data analyses (dashed line)
\cite{isupov-web,blin19,mok20p}. The systematic uncertainties on the data are shown by the shadowed areas at the bottom
of each plot.}
\label{fig:LTprime_05_3}
\end{figure}

Determination of $\sigma_{LT'}$ was made through the BSA $A_{LT'}$ and a parameterization of the $\sigma_{0}$ cross sections from the 
previous measurement ~\cite{markov20}:
\begin{eqnarray}
A_{LT'} 
&=&
\frac{\sqrt{2\epsilon_L(1-\epsilon)}\,\sigma_{LT'}\,\sin\,\theta_{\pi}\,\sin\,\phi_{\pi}}{\sigma_{0}}.
\label{eq:altp}
\end{eqnarray}
The $A_{LT'}$ values are multiplied by $\sigma_0$ and the structure function $\sigma_{LT'}$ was then extracted using Eq.(\ref{eq:altp}) through 
fitting the $\phi_{\pi}$ distributions in each bin of $(Q^2,W,\cos\theta_{\pi})$. 
The systematic uncertainties of $\sigma_{LT'}$ are less than $10\%$ and arise mainly from the uncertainties of the beam polarization $P_e$ and 
the uncertainties of $\sigma_0$ from the available measurements~\cite{markov20}. The contributions from the systematic uncertainties for the BSAs 
are much smaller and were not included in the evaluation of the systematic uncertainties for $\sigma_{LT'}$. Representative examples of the structure 
functions $\sigma_{LT'}$ at $W > 1.6$~GeV are shown in Fig.~\ref{fig:LTprime_05_3}. The full set of our results for $\sigma_{LT'}$ can be found in
Ref.~\cite{clasdb}.

Results on the $\sigma_{LT'}$ structure function for $\pi^0p$ electroproduction in the third resonance region are presented here for the first
time. In Fig.~\ref{fig:LTprime_05_3} we also show the comparison between the data on $\sigma_{LT'}$ from our measurements and the expectations 
from MAID~\cite{MAID}. This unitarized reaction model incorporates all well-established resonances parameterized using Breit-Wigner amplitudes 
and with backgrounds calculated from $t$-channel vector-meson exchange and other Born diagrams. 
The $\sigma_{LT'}$ are computed with the $\gamma_{v}pN^*$ resonance electro\-couplings obtained from the analyses of the $\pi N$ electro\-production channels only \cite{Tiator:2011pw} and from the CLAS $\pi N$ and $\pi^+\pi^-p$ electro\-production data \cite{isupov-web,blin19,mok20p}.  

\begin{figure*}[htb]
\includegraphics[scale=0.8]{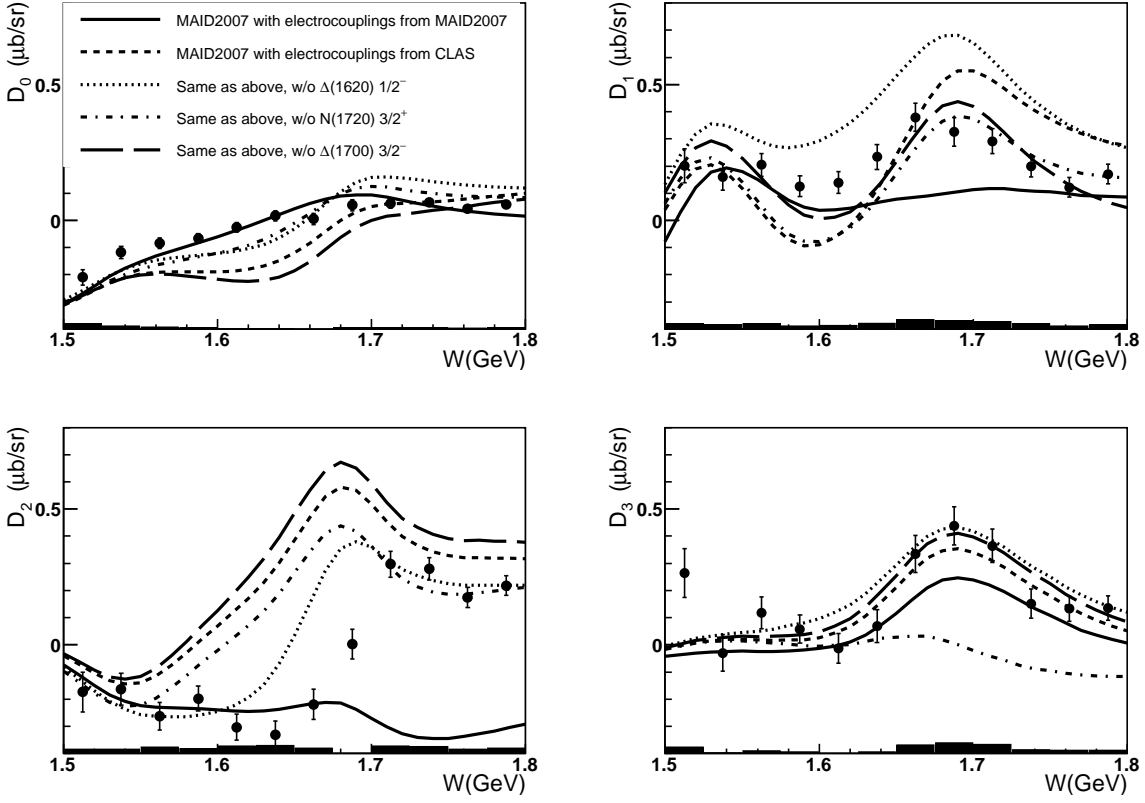}
\caption{Legendre moments $D_l(Q^2,W)$ ($l$=0,1,2,3) of the $\sigma_{LT'}$ structure function from the $\pi^0p$ electroproduction data at
$Q^2$=0.4-0.6~GeV$^2$: $D_0(Q^2,W)$ (top left), $D_1(Q^2,W)$ (top right), $D_2(Q^2,W)$ (bottom left), $D_3(Q^2,W)$ (bottom right). The experimental 
results are shown by the filled circles with error bars accounting for the statistical uncertainties. The systematic data uncertainties are shown in 
the bottom part of each plot. The computed $D_l$ moments within the MAID model~\cite{MAID} with the $\gamma_{v}pN^*$ electrocouplings from 
Ref.~\cite{Tiator:2011pw} and from Refs.~\cite{isupov-web,blin19,mok20p} are shown by the thick solid and dashed lines, respectively. We also show 
the computed $D_l$ values within the MAID model~\cite{MAID} with the $\gamma_{v}pN^*$ electrocouplings from Refs.~\cite{isupov-web,blin19,mok20p} 
when the contributions from particular resonances are taken out: $\Delta(1620)1/2^-$ (thin dotted lines), $\Delta(1700)3/2^-$ (thick long-dashed lines), $N(1720)3/2^+$ (thin dash-dotted lines).}
\label{fig:D0123}
\end{figure*}

In order to explore the sensitivity of $\sigma_{LT'}$ to the contributions from nucleon resonances ($N^*$), we
decomposed this structure function using a Legendre polynomial expansion for each bin of $(Q^2,W)$ to determine 
the Legendre moments $D_l(Q^2,W)$:
\begin{align}
\sigma_{LT'} &= \sum_{\ell = 0}^{\ell_{\rm max}}D_{\ell}P_{\ell}(\cos\theta_{\pi}),\label{legD}
\end{align}
with $l$ from 0 to 3. The expansion of Eq.(\ref{legD}) is truncated at $l_{max}=3$ to provide a stable description of the $\cos\theta_{\pi}$ dependence 
of $\sigma_{LT'}$. The results on the $W$ dependence of the Legendre moments $D_{l}$ are shown in Fig.~\ref{fig:D0123}. The full set of extracted 
Legendre moments is available in Ref.~\cite{clasdb}.
\noindent
\noindent
In order to relate the Legendre moments $D_{l}$ to the bilinear products of the $\pi^0p$ multipole electroproduction amplitudes, the formalism 
developed in Ref.~\cite{tiator95} is used. The $\sigma_{LT'}$ structure function is expressed in terms of bilinear combinations of the $F_{i}$ 
CGLN amplitudes, as described in Appendix C of Ref.~\cite{tiator95}. Eqs.~(23-28) in that paper allow us to relate the CGLN amplitudes $F_{i}$
($i$=1,..,6) to the bilinear products of the multipole amplitudes that enter into the $D_l$ \mbox{Legendre} moments. Since the bilinear product of
multipoles in the \mbox{Legendre} moments $D_l$ of different $l$-values contain the contributions from resonances of different spins and parities, the 
\mbox{Legendre} moments $D_l$ are suitable for disentangling the electroexcitation of different nucleon resonances.




We explored the sensitivity of the $D_l$ Legendre moments to the variation of the $\gamma_{v}pN^*$ electrocouplings of all pronounced resonances 
in the second and third resonance regions and, in particular, to the $\Delta(1620)1/2^-$, $\Delta(1700)3/2^-$, and $N(1720)3/2^+$ electrocouplings that
currently have been established from solely the $\pi^+\pi^-p$ electroproduction data. 
The sensitivity of the $D_l$ moments to the variation of the electrocouplings of all prominent resonances can be demonstrated by computing them 
within the MAID model~\cite{MAID} with the $\gamma_{v}pN^*$ electrocouplings from only the $\pi N$ electroproduction data~\cite{Tiator:2011pw} and 
from the CLAS results~\cite{isupov-web,blin19,mok20p} on the $\gamma_{v}pN^*$ electrocouplings determined from both $\pi N$ and $\pi^+\pi^-p$ 
electroproduction as shown by the thick solid and dashed lines in Fig.~\ref{fig:D0123}, respectively. The sensitivity of the $D_l$ moments to the contribution from each of the states $\Delta(1620)1/2^-$,
$\Delta(1700)3/2^-$ and $N(1720)3/2^+$ was studied by turning off each in turn while taking the 
$\gamma_v pN^*$ electrocouplings of the other resonances from the CLAS results ~\cite{isupov-web,blin19,mok20p}. 
The sensitivity to the particular resonance contributions can be seen in Fig.~\ref{fig:D0123} as the difference between the computed $D_{l}$ moments 
when the contributions from  all resonances are taken into account (thick dashed lines) and when the contribution from a particular resonance is 
turned off.  
All $D_l$ moments demonstrate variations outside the data uncertainties when the $\gamma_{v}pN^*$ electrocouplings from $\pi N$ electroproduction
\cite{Tiator:2011pw} are replaced by the CLAS results~\cite{isupov-web,blin19,mok20p}, suggesting sensitivity of the $\sigma_{LT'}$ data to the 
resonance contributions in these kinematics.

The $D_{0}$ moment demonstrates sensitivity to the $\Delta(1620)1/2^-$ resonance (see Fig.~\ref{fig:D0123} top left). This sensitivity is due to 
the multipole contributions of the $D_{0}$ decomposition:  
\begin{eqnarray}
D_0 \sim (5E^*_{3+}-2E^*_{3-}+M^*_{1-}+M^*_{1+})S_{0+}\\ \nonumber
+E^*_{0+}(S_{3-}-S_{3+}).
\label{s31d0mult}
\end{eqnarray}
We are using the well-known notations for the multipoles explained in Ref.~\cite{tiator95}. The impact from the $\Delta(1620)1/2^-$ on the $D_{0}$ 
moment emerges from the interference between the $S_{0+}$ resonance multipole with the transverse multipoles from the non-resonant contributions, as 
well as from interference between the resonance $E_{0+}$ transverse and longitudinal multipoles with the non-resonant contributions. 

From the $D_{1}$ multipole decomposition, we found that $D_1$ should be sensitive, in particular, to the contributions from the $\Delta(1700)3/2^-$ 
and $N(1720)3/2^+$ resonances since the $D_1$ moment contains the products:
\begin{eqnarray}
D_1 \sim -6E^*_{2-}S_{2-}-6M^*_{2-}S_{2-}\\
+6E^*_{1+}S_{1+}-6M^*_{1+}S_{1+},
\label{d33mult}
\nonumber
\end{eqnarray}
where the multipole and the multipole conjugated products in the first two terms contain contributions from the $\Delta(1700)3/2^-$ and in the 
second two terms from the $N(1720)3/2^+$. The expected sensitivities are supported by the data on the $D_1$ moment (see Fig.~\ref{fig:D0123} top right). 

At $W$ from 1.65~GeV to 1.70~GeV, the $D_{2}$ moment evolves rapidly and changes sign (see Fig.~\ref{fig:D0123} bottom left). Making use of the CLAS 
results on the $\gamma_{v}pN^*$ electrocouplings~\cite{isupov-web,blin19,mok20p} allows us to reproduce this trend even when the contributions from 
either the $\Delta(1620)1/2^-$, $\Delta(1700)3/2^-$, or $N(1720)3/2^+$ are turned off. 

The $D_{2}$ moment is sensitive to the $\Delta(1620)1/2^-$ and $N(1720)3/2^+$ resonances.  This sensitivity emerges from the following terms in the 
$D_2$ multipole decomposition: 
\begin{eqnarray}
D_2 \sim 12(M^*_{2+}-E^*_{2-})S_{1+}\\ \nonumber
+6(3E^*_{2+}+2M^*_{2+})S_{1+}-15M^*_{1+}S_{2-} \\ \nonumber
+5(5E^*_{3+}-2E^*_{3-}+M^*_{3-}-M^*_{3+})S_{0+}\\ \nonumber
+5E^*_{0+}(3S_{3-}-4S_{3+}).
\label{p13bmult}
\end{eqnarray} 
The $N(1720)3/2^+$ resonance impacts the $D_2$ moments owing to the interference between its longitudinal $S_{1+}$ multipole and the transverse 
multipoles from the non-resonant contributions, as well as from  the interference of the $M_{1+}$ transverse resonance multipole and the 
longitudinal part of the non-resonant contributions. The $\Delta(1620)1/2^-$ impacts the $D_2$ moment because of the interference between its 
$S_{0+}$ multipole and transverse non-resonant contributions, as well as due to the interference between the transverse $E_{0+}$ multipole of 
the resonance and the longitudinal parts of the non-resonant contributions.

The $D_3$ moment is strongly affected by the $N(1720)3/2^+$ resonance (see Fig.~\ref{fig:D0123} bottom right) due to the interference between the 
$S_{1+}$ multipole for the resonance and the transverse part of the non-resonant processes, as well as because of the interference between the 
transverse $M_{1+}$ resonance multipole and the longitudinal part of the non-resonant contribution seen in the multipole decomposition of $D_3$:
\begin{eqnarray}
D_3 \sim 18(M^*_{3-}-E^*_{3+})S_{1+}\\ \nonumber
+(34E^*_{3+}-36E^*_{3-})S_{1+}
-28M^*_{1+}S_{3+}.
\label{p13bd3mult}
\end{eqnarray}
Studies of the Legendre moment sensitivities to the resonant contributions at 0.6~GeV$^2 < Q^2 < 1.0$~GeV$^2$ revealed similar 
features as those observed for the data at 0.4~GeV$^2 < Q^2 < 0.6$~GeV$^2$ and discussed above.

In summary, the beam spin asymmetries and the $\sigma_{LT'}$ structure functions have become available from the CLAS data on exclusive $\pi^0p$
electroproduction at 1.5~GeV $< W < 1.8$~GeV and photon virtualities 0.4~GeV$^2 < Q^2 < 1.0$~GeV$^2$. These observables for $\pi^0p$ electroproduction 
off protons were obtained for the first time at the invariant masses of the final state hadrons $W > 1.6$~GeV. The Legendre moments $D_l$ ($l$=0,1,2,3) 
were determined from fits of the angular dependencies of $\sigma_{LT'}$ in each bin of $W$ and $Q^2$ covered by the measurements. 
The $D_0$ and $D_1$ Legendre moments demonstrate sensitivity to the electroexcitation amplitudes of the $\Delta(1620)1/2^-$, $\Delta(1700)3/2^-$, 
and $N(1720)3/2^+$ resonances. Previously, the information on the electroexcitation amplitudes of these excited states was obtained in the studies 
of $\pi^+\pi^-p$ electroproduction off protons with CLAS \cite{Mokeev20,Mok20,mok20p}. Combined studies of the exclusive $\pi^+n$ and $\pi^0p$
electroproduction channels will provide independent results on the electroexcitation amplitudes of the nucleon resonances in the mass range of 
$W > 1.6$~GeV. Comparison of the results on the resonance electroexcitation amplitudes from the studies of both $\pi N$ and $\pi^+\pi^-p$ 
electroproduction will allow us to shed light on the systematic uncertainties in their extraction. 


We acknowledge the outstanding efforts of the staff of the Accelerator and the Physics Divisions at Jefferson Lab in making this experiment possible. 
This work was supported in part by the U.S. Department of Energy, the National Science Foundation (NSF), the Italian Istituto Nazionale di Fisica Nucleare (INFN), 
the French Centre National de la Recherche Scientifique (CNRS), the Skobeltsyn Institute of Nuclear Physics, the Physics Department at Moscow State University, 
the French Commissariat pour l$^{\prime}$Energie Atomique, the UK Science and Technology Facilities Council, and the National Research Foundation (NRF) of Korea. 
The Southeastern Universities Research Association (SURA) operates the Thomas Jefferson National Accelerator Facility for the U.S. Department of Energy 
under Contract No. DE-AC05-06OR23177. This work was supported in part by the Chilean National Agency of Research and Development ANID PIA/APOYO AFB180002. 
The work is also supported in part by DOE grant DE-FG02-04ER41309.


\end{document}